\documentclass[a4paper, twocolumn, 11pt]{IEEEtran}  

\IEEEoverridecommandlockouts                              

\usepackage{mathptmx} 
\usepackage{times} 
\usepackage{amsmath} 
\usepackage{amssymb}  
\usepackage{color, soul}	
\usepackage{multirow}
\usepackage{array}
\usepackage{booktabs}
\usepackage{fancyhdr} 

\newcommand{\refeqn}[1]{(\ref{#1})}

\newcommand{\RT}{}

\newcommand{\RTT}{}

\usepackage[pdftex]{graphicx}
\usepackage{epstopdf} 
\graphicspath{{CalorimeterPaper/Figures/}}
\DeclareGraphicsExtensions{.pdf,.jpeg,.png,.eps,.jpg}

\usepackage{todonotes}

\begin{document}
\title{\RT High Speed Peltier Calorimeter for the Calibration of High Bandwidth Power Measurement Equipment}

\author{Damien~F.~Frost~\IEEEmembership{Student Member,~IEEE},~David~A.~Howey~\IEEEmembership{Member,~IEEE}%
\thanks{D.A.~Howey and D.F.~Frost are based in the Energy and Power Group, Department of Engineering Science of the University of Oxford, Parks Road, Oxford, OX1 3PJ, United Kingdom. Emails: {\tt\small \{damien.frost, david.howey\} at eng.ox.ac.uk}.
}
}

\maketitle

\begin{abstract}

Accurate power measurements of electronic components operating at high frequencies are vital in determining where power losses occur in a system such as a power converter. Such power measurements must be carried out with equipment that can accurately measure real power at high frequency. We present the design of a high speed calorimeter to address this requirement, capable of reaching a steady state in less than 10 minutes. The system uses Peltier thermoelectric coolers to remove heat generated in a load resistance, and was calibrated against known real power measurements using an artificial neural network. A dead zone controller was used to achieve stable power measurements. The calibration was validated and shown to have an absolute accuracy of $\pm$8 mW (95\% confidence interval) for measurements of real power from 0.1 to 5 W. 

\end{abstract}


\section{Introduction}\label{sec:Introduction}

Energy efficiency is crucial for optimum thermal and {\RTT electrical performance in modern electronics} such as switching power converters. In order to obtain accurate efficiency estimates of a system or component, accurate measurements of the voltage and current waveforms of a device under test (DUT) must be acquired. However, such measurements are sometimes difficult to acquire accurately if the DUT has very fast current and/or voltage transients. For example, the voltage and current waveforms in high frequency magnetic components in a switched mode power supply include high frequency harmonics that may contribute to real power loss. Devices which exhibit fast transients may be tested using precision power analyzers (PPAs) capable of detecting high frequency voltage and current waveforms in the MHz range. {\RT In order to calibrate PPAs, this paper presents a new technique and novel high speed calorimeter to measure the real power output as heat.}

Calorimetry has a long history and is typically used to determine the amount of energy released or absorbed in chemical reactions \cite{5669610}. More recently, it has been used to accurately determine the efficiency of motors \cite{790935}, power converters \cite{5544503}, consumer electronic devices \cite{4528875}, power electronic devices \cite{chen2002apparatus, 6656219, 4215024}, lithium-ion batteries \cite{chen2014measurements, Feng2014294}, and ferromagnetic materials \cite{5004697}. {\RTT In radio frequency applications, calorimetry has been used to measure the incident power into radio receivers, and to calibrate radio frequency equipment \cite{481327,4213147,fantom1990radio}.}

{\RT Calorimetry is an attractive technique for measurement of real power because the voltage and current waveforms of the DUT have relatively small effect on measurement accuracy.} However, in some cases a single measurement can take up to 12 hours to obtain \cite{5669610}, because calorimetric measurements must be taken at thermal steady state and it may take considerable time to reach a thermal equilibrium, depending on the size of the DUT and the characteristics of the calorimeter. {\RT A calorimeter which can produce a reading in a short amount of time, could be integrated into the production cycle of measurement equipment such as PPAs.} {\RT Calorimeters designed to measure radio frequency power have been reported that are capable of reaching a measurement within 10 minutes \cite{6375844}, however they have a very limited measurement range: 0.7 mW is the largest power measured in \cite{6375844}. We present a calorimeter that can reach thermal equilibirum in less than 10 minutes , and can measure power up to 5 W.}

\section{Operating Principle and System Design}\label{sec:OpPrin}

In the calorimeter presented in this paper, the DUT is specifically designed to have a small size and a flat frequency response. The real power dissipated by the DUT is compared to the real power calculated by the PPA. The calibration process of a PPA with the calorimeter can be summarized as follows:

\begin{enumerate}
\item Calibrate the calorimeter using the PPA by driving the DUT in the calorimeter with a DC or low frequency voltage source (since the calibration of the PPA is known to be accurate in this region).
\item Drive the same DUT with a high frequency voltage and current while measuring its heat output directly, and its voltage and current with the PPA to be calibrated.
\item Compare the calibrated heat output from the calorimeter to the real power reading of the PPA to produce a calibration relationship.
\end{enumerate}

In this calibration technique, it is assumed that the low frequency and high frequency calorimeter power measurements are equivalent, which is a reasonable assumption since we are measuring the heat output of the DUT, and has been shown in the past to be a valid calibration technique \cite{5339204}.

\begin{figure}
\centering
\includegraphics[width=\linewidth]{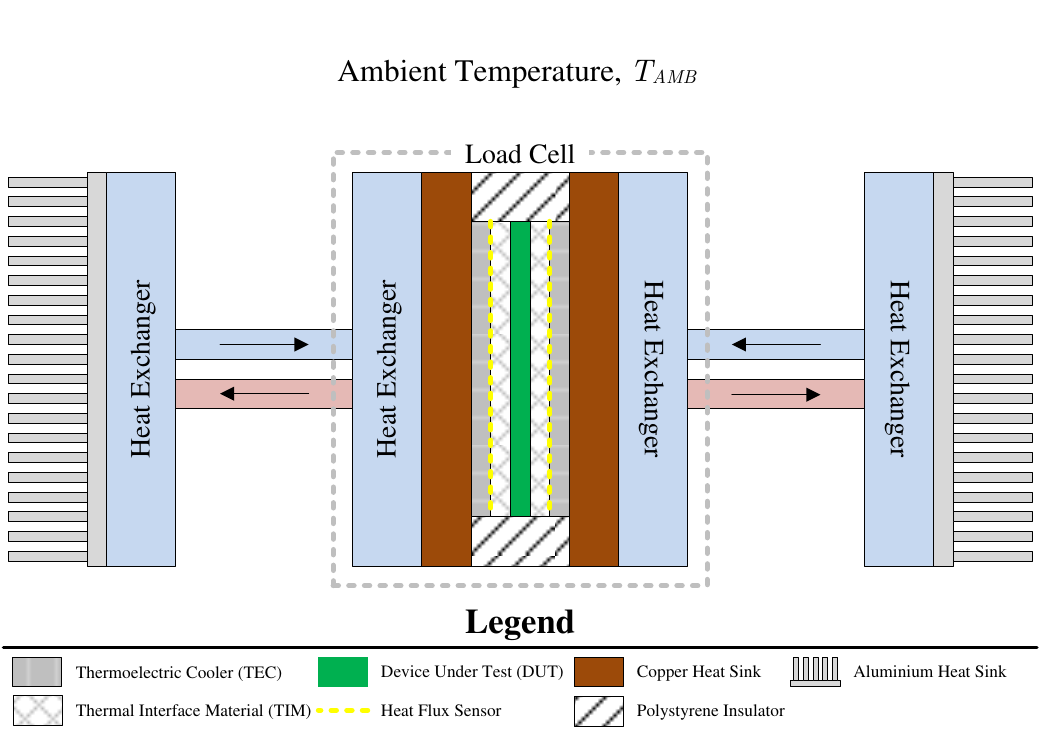}
\caption{Schematic diagram of the proposed calorimeter showing a cutaway of the Load Cell.}
\label{fig:CalorimeterDiagram}
\end{figure}

The proposed calorimeter (Fig.\ \ref{fig:CalorimeterDiagram}) is based on a small `sandwich' style calorimeter \cite{5004697}. This type of calorimeter uses conduction to remove the heat generated by the DUT and therefore reaches thermal steady state rapidly since it has a small thermal mass and low thermal resistance down the main heat removal path.
 
The term \emph{load cell} will be used here to describe the testing area of the calorimeter, comprising the DUT, thermal interface material, heat flux sensors, peltier coolers and heat sinks, as shown in the centre of Fig.\ \ref{fig:CalorimeterDiagram}. At the centre of the load cell is the DUT comprising a high frequency resistive load (HFL). Unlike traditional calorimeters, this is an integral and fixed part of the system and is designed not to be removed. This is surrounded by a thermal interface material to ensure a good conduction path to the two thermoelectric coolers (TECs) on either side. These control the heat flux from the DUT. Heat transferred through and produced by the TECs is removed from the calorimeter through water based heat exchangers. The entire load cell can be enclosed in a temperature regulated chamber, if necessary.

The power loss from the DUT may be measured using two methods: indirectly using the characteristics of the TECs, or directly using the embedded heat flux sensors shown in Fig.\ \ref{fig:CalorimeterDiagram}. %

In the ideal case, all of the heat generated by the DUT is absorbed through the thermal interface material and the TECs, and therefore corresponds directly to the heat measured by both heat flux sensors. In practice, this is not the case, since there exists small and unavoidable thermal leakage paths that cause measurement errors and must be accounted for through calibration as discussed below.

\subsection{High Frequency Load Design}
\begin{figure}[h]
\centering
\includegraphics[width=0.65\linewidth]{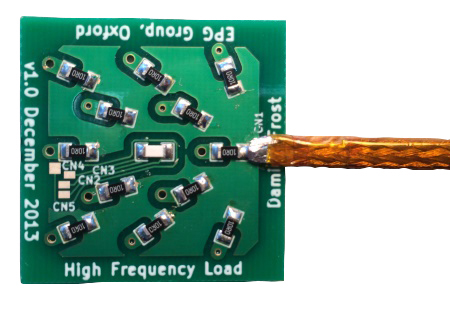}
\caption{High frequency load (HFL) with the power leads attached. The white resistor at the centre is a platinum resistor thermometer, all others are load resistors.}
\label{fig:HFL}
\end{figure}

It was desired that the DUT should behave as an ideal resistor for the largest bandwidth as possible, thus it was designed with minimal inductance and capacitance. In order to achieve this, 12 surface mount resistors were arranged on either side of a 35 mm x 35 mm two-layer printed circuit board (PCB) (Fig. \ref{fig:HFL} ) in such a way that the magnetic fields generated by the high frequency currents are substantially cancelled out. The resistors were spread evenly around the PCB to ensure even heat generation. A platinum resistance thermometer (PRT) temperature sensor was mounted in the centre on either side of the PCB.

\subsection{Thermal Simulation}\label{sec:thermalsim}
Prior to construction, a finite element numerical simulation of the calorimeter was undertaken using a 3D model in Autodesk Simulation 2013 to quantify the expected performance. In order to reduce computation time, the following assumptions were made:
\begin{itemize}
\item The heat sinks of the load cell were modelled as copper blocks each having an isothermal surface.
\item The thermal impact of the wires required to power the load was ignored.
\item The ambient temperature was held at a constant 25 $^{\circ}\mathrm{C}$.
\end{itemize}
	
{\RT The heat generation was modelled as a dc heat source, the same size as the HFL PCB. }The heat flux through the TECs could be directly calculated in the simulation. Comparing this value to the heat generated by the simulated DUT gives an indication of the expected measurement accuracy of the system under the above simplifying assumptions.
 
 \begin{figure}
 \centering
 \includegraphics[width=0.7\linewidth]{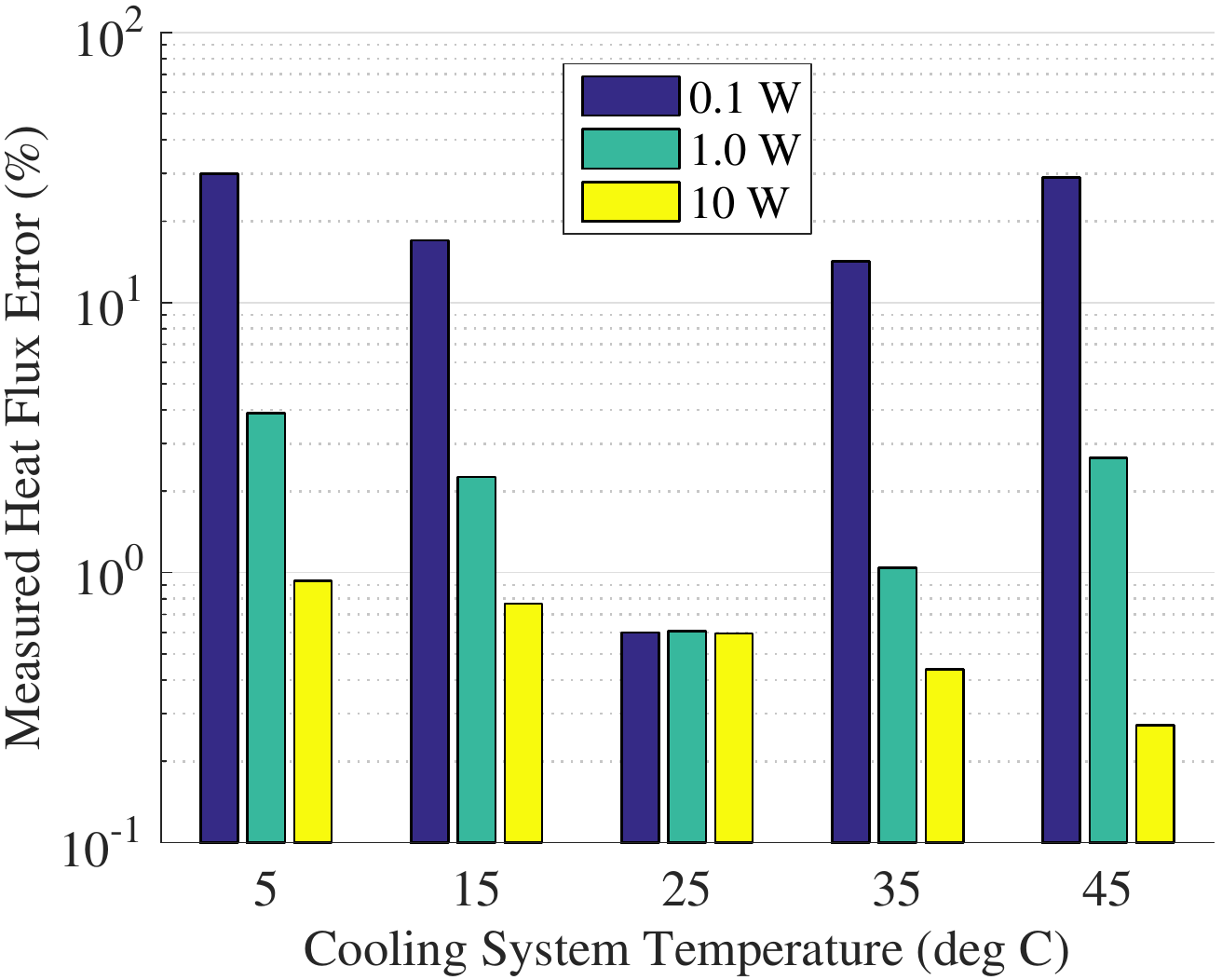}
 \caption{Simulated relative measurement error as a function of cooling system temperature (i.e.\ temperature of DUT and heat sinks) at various levels of heat generation.}
 \label{fig:TSimVaryCoolSys}
 \end{figure}
 
 \begin{figure}
  \centering
  \includegraphics[width=0.7\linewidth]{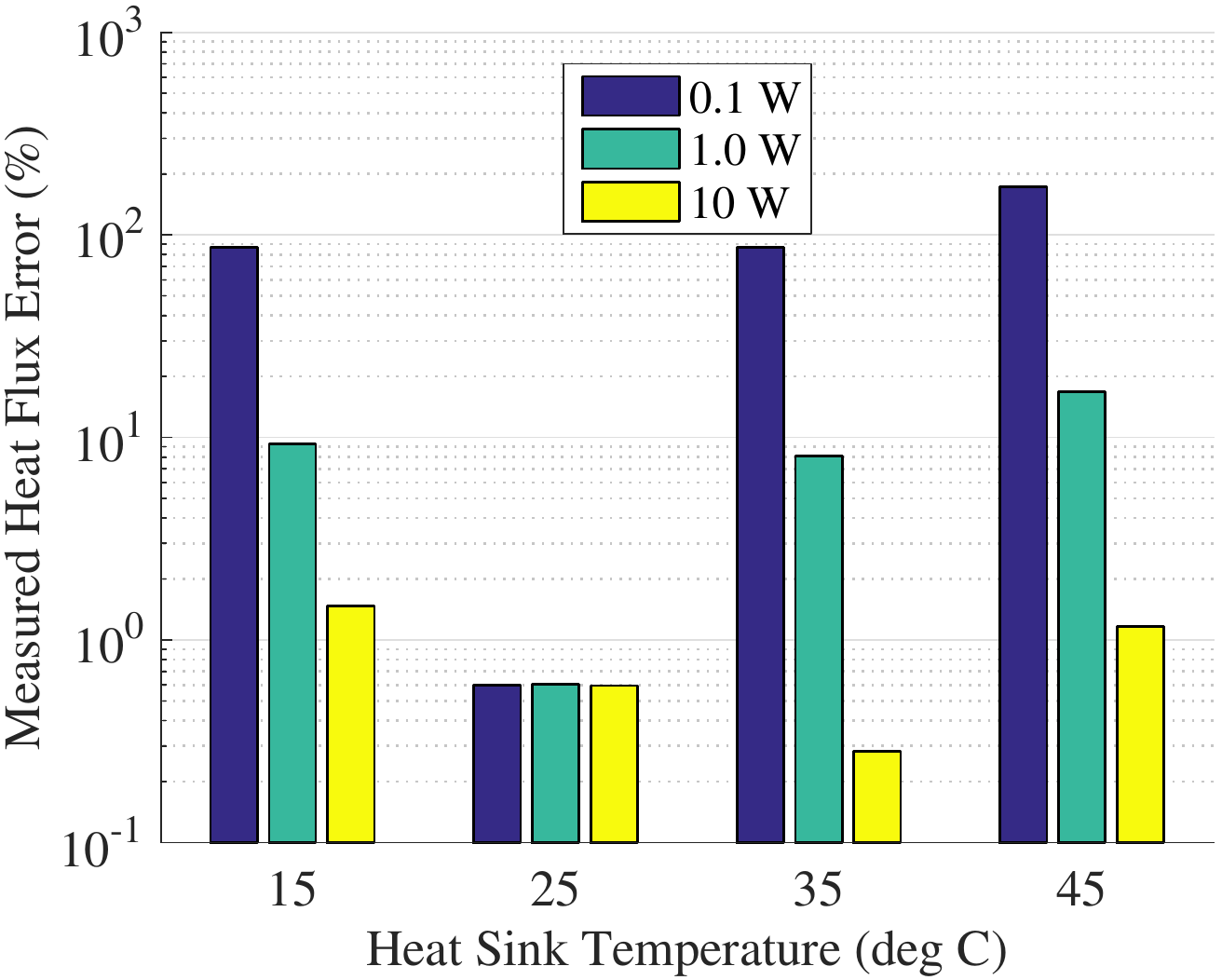}
  \caption{Simulated relative measurement error as a function of heat sink temperature. Note that the error for the $0.1$ W case is much greater than 30\% for heat sink temperatures other than 25 $^{\circ}\mathrm{C}$.}
  \label{fig:TSimVaryHeatSink}
  \end{figure}
	 
Fig.\ \ref{fig:TSimVaryCoolSys} shows the simulation results when the cooling system temperature is varied. The cooling system temperature is defined as the temperature that the DUT is controlled to and the temperature of the heat sinks; in this set of simulations these were both set to the same temperature. The cooling system temperature was varied between 5 $^{\circ}\mathrm{C}$ and 45 $\,^{\circ}\mathrm{C}$ in 10 $^{\circ}\mathrm{C}$ increments. The ambient temperature was set to 25 $^{\circ}\mathrm{C}$. The relative percent error was calculated with \refeqn{error}:
\begin{equation}
	\label{error}
	e_{meas}=\left| \frac{\left( q_{meas}-P_{DUT} \right)}{P_{DUT}} \right| 100 {\RT \%}
\end{equation}
where $e_{meas}$ is the relative error in the power measurement, $q_{meas}$ is the total measured heat flux out of the DUT, and $P_{DUT}$ is the power applied to the DUT.

As expected, the smallest error occurs when the TEC control temperature matches the ambient temperature at 25 $^{\circ}\mathrm{C}$, since under this scenario the temperature gradient between the ambient temperature and DUT is theoretically zero. At this temperature, the error is 0.6\% for all three power levels simulated. However, the error between the power levels changes dramatically as the TEC control temperature varies. 

\begin{figure}
\centering
\includegraphics[width=\linewidth]{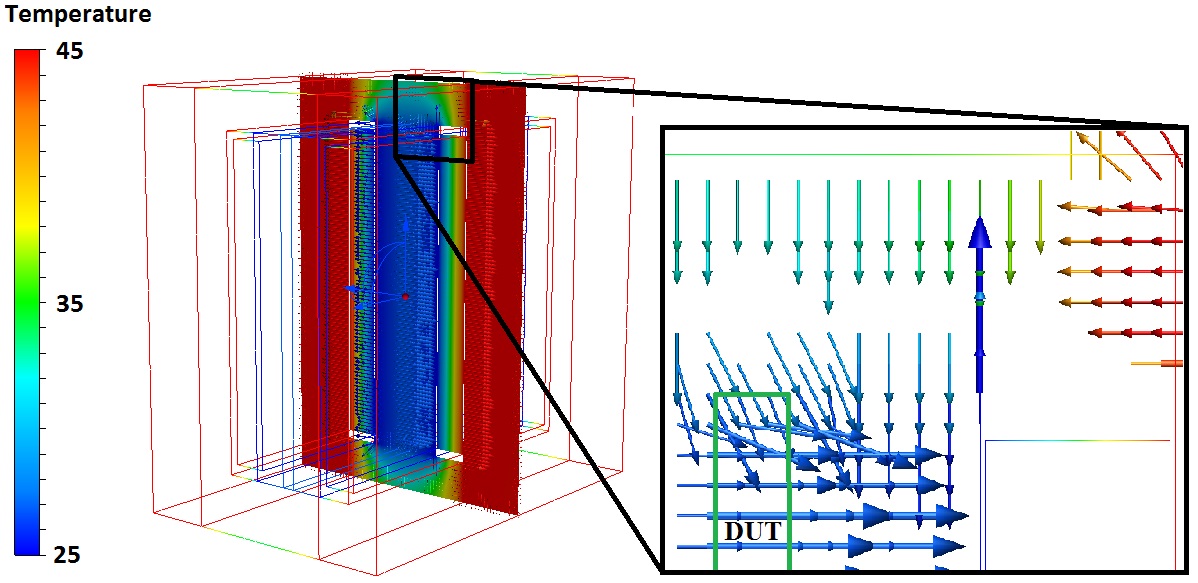}
\caption{Temperature and heat flux visualization from the finite element simulations of the calorimeter. The heat sink temperature was regulated to 45 $^{\circ}\mathrm{C}$, the DUT 25 $^{\circ}\mathrm{C}$, and the ambient temperature 25 $^{\circ}\mathrm{C}$. The nominal power of the DUT was 1 W. The length of the arrows is proportional to the heat flux at the boundary of each part of the load cell.}
\label{fig:45degHSTotal}
\end{figure}

Fig.\ \ref{fig:TSimVaryHeatSink} shows the simulation results when the temperature of the heat sinks was varied while the TEC control temperature and the ambient temperature, $T_{AMB}$, remained constant at 25 $^{\circ}\mathrm{C}$. As shown, the relative error also increases between the power levels as the heat sink temperature varies. In this case, the effect is more dramatic, and at a heat sink temperature of 45 $^{\circ}\mathrm{C}$, the error in the 0.1 W case is 172\%.

Fig.\ \ref{fig:45degHSTotal} shows a heat flux and temperature visualization through a plane of the simulated load cell, when the heat sink temperature was 45 $\,^{\circ}\mathrm{C}$ and the TEC control temperature and the ambient temperature were 25 $\,^{\circ}\mathrm{C}$. As shown in the inset, there is a heat leakage path originating from the heat sinks, into the insulation barrier, and finally into the DUT. This additional heat flux is measured by the heat flux sensor, producing the erroneous measurements seen in the previous figures. This effect was addressed through calibration, as discussed in Section \ref{calibration}.

\section{Hardware}
The frame of the load cell and many other custom parts were designed with Autodesk Inventor, and manufactured with a consumer grade 3D printer (Ultimaker Original). This allowed the load cell to have an  intricate design including areas to route sensing wires and features to mount sensors, TECs, the thermal interface material, and insulation.

The 3D printing material used was poly-lactic  acid (PLA) which has a glass transition temperature between 60 - 65 $\,^{\circ}\mathrm{C}$ \cite{Sodergard20021123}. As a consequence, the working temperature of the calorimeter using this manufacturing technique is limited to about 50 $\,^{\circ}\mathrm{C}$. Careful attention was taken to use non-ferromagnetic materials throughout the calorimeter to ensure that the only real power consumed was due to the HFL within the load cell.

\begin{figure}
\centering
\includegraphics[width=\linewidth, trim=0.9cm 0cm 0cm 0cm]{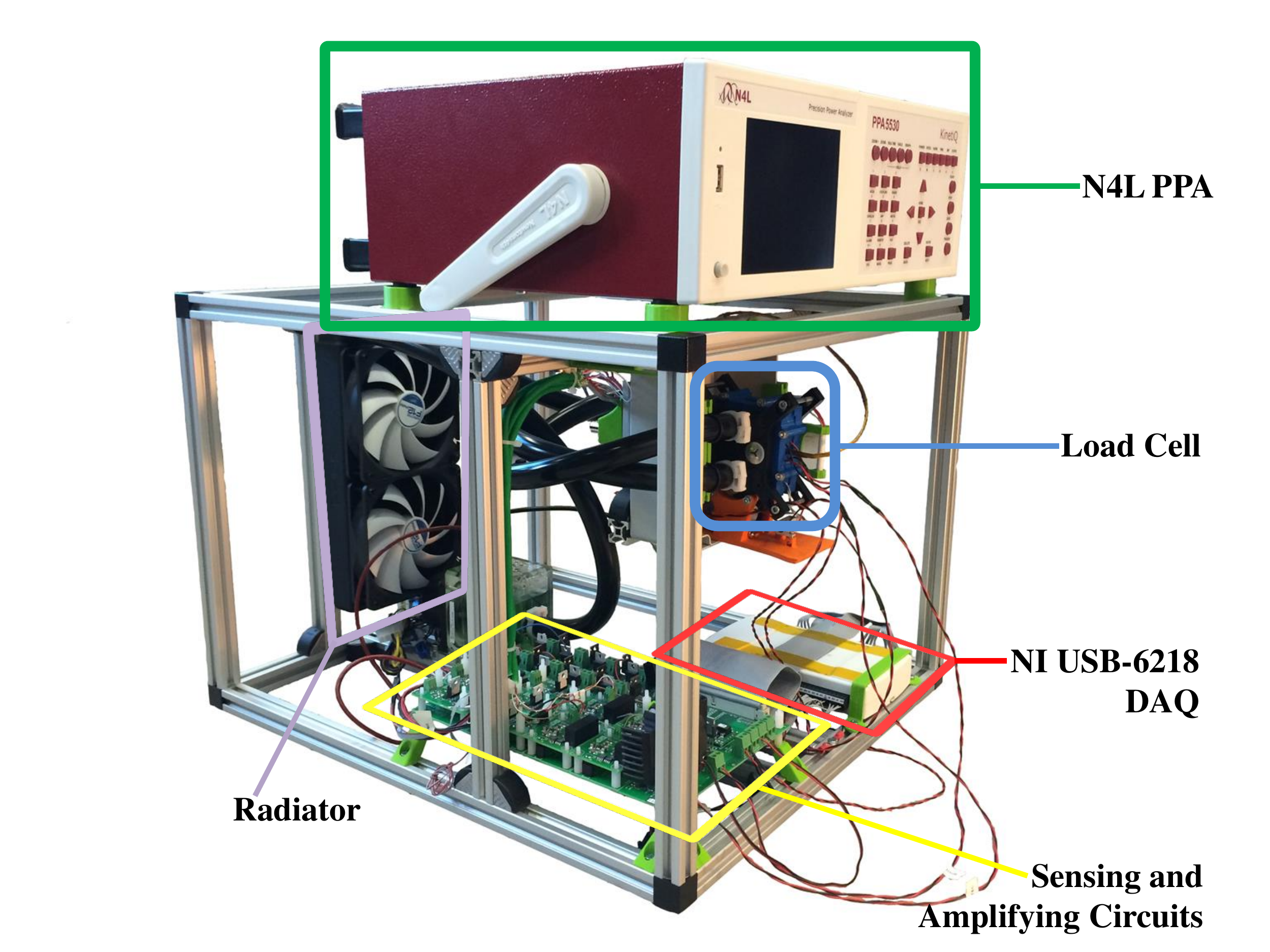}
\caption{Photo of the completed calorimeter. The overall dimensions excluding PPA are 440 x 500 x 360 mm.}
\label{fig:ExperimentalSetup}
\end{figure}

A photo of the complete calorimeter is shown in Fig.\ \ref{fig:ExperimentalSetup}, with the {\RT Newton's 4th Precision Power Analyzer model 5530 (N4L PPA)} mounted above. The system interfaced to a computer through a National Instruments USB-6218 data acquisition board (NI DAQ). A graphical user interface was developed for control and the taking of measurements.

The TECs used in the calorimeter were manufactured by Multicomp, part number MCS-127-10-25-S. The heat flux sensors used were Omega type HSF-4, {\RTT which  were individually calibrated by Omega to a sensitivity of 2.0 $\mu \mathrm{V}/\mathrm{W}/\mathrm{m}^2$ and were connected to the NI DAQ via an amplifier.} The PCB temperature sensors were from Innovative Sensor Technology part number P0K1.1206.2P.A, {\RTT and are rated to Class A of the IEC 60751 specification, $\pm$(0.15 $+$ 0.002$\left|\mathrm{T}\right|$).} The ambient temperature sensor was from Labfacility, part number 010010TD, and the water temperature sensor was Labfacility, part number XE-3630-001. {\RTT Both of the Labfacility temperature sensors are rated to Class B of the IEC 60751 specification, $\pm$(0.3 $+$ 0.005$\left|\mathrm{T}\right|$). All three types of temperature sensors were connected to the NI DAQ via current transmitters.}

\section{Control}
\begin{figure}
\centering
\includegraphics[width=\linewidth]{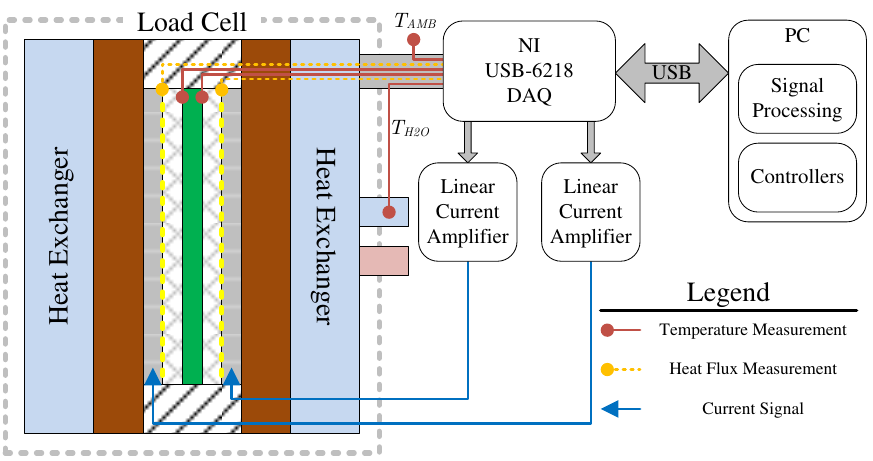}
\caption{Schematic diagram of the sensing and control system used in the calorimeter.}
\label{fig:CalControl}
\end{figure}

A simplified diagram of the sensing and control systems used in the calorimeter is shown in Fig.\ \ref{fig:CalControl}.  There are six sensors used for the calorimetric measurement: two PRT temperature sensors on either side of the DUT, two heat flux sensors, a PRT water temperature sensor, and a PRT ambient temperature sensor. Temperature inside the load cell is regulated by controlling the current through the two TECs through two custom built linear current amplifiers.

\subsection{TEC Control}\label{sec:TECCtrl}
The TECs provide relatively fast response of heat flux to changes in driving current. However, the amount of heat they can pump is very closely related to the temperature gradient across the device, and this can introduce unwanted perturbations in heat flux. 

{\RT Therefore, the worst case sensitivity in the change in TEC heat pumping power, $\Delta W_{pump}$, was modeled.}
Using information obtained from the TEC datasheet \cite{MCS1271025S}, a curve representing the change in heat pumping power for a given temperature change across the device may be fitted; \refeqn{eqn:TECdW} gives the curve found for the TECs used in the calorimeter:
\begin{equation}
\frac{\Delta W_{pump}}{\Delta T} = 0.051 \cdot I_{TEC} + 0.169; \textrm{ } I_{TEC}\leq 2.0 \text{ A} \label{eqn:TECdW}
\end{equation}
where $\Delta T$ is the change in temperature across the TEC, and $I_{TEC}$ is the current in the TEC.

As an example, when $I_{TEC}$ = 1 A, a 1 $^{\circ}\mathrm{C}$ increase in the temperature across the TEC will result in a 220 mW decrease in heat being pumped through the TEC. Therefore, the coolant system must be designed to maintain a stable coolant temperature, and closed loop feedback control must be used to reduce the sensitivity of the system to changes in coolant temperature. Additionally it can be noted that the pumping power is sensitive to changes in the driving current $I_{TEC}$, particularly when the temperature across the TEC is close to 0 $\,^{\circ}\mathrm{C}$ and $I_{TEC}$ is close to zero, which occurs when the calorimeter is measuring very low powers. Using the information from the TEC datasheet, one can fit another heat pumping sensitivity equation relating changes in driving current to changes in heat pumping power, \refeqn{eqn:Wpump}:
\begin{equation}
\frac{\Delta W_{pump}}{\Delta I_{TEC}} = -7.41 \cdot I_{TEC} + 17.23; \textrm{ } I_{TEC}\leq 2.0 \text{ A} \label{eqn:Wpump}
\end{equation}
where $\Delta I_{TEC}$ is the change in current through the TEC.

Therefore, in the worst case scenario the heat pumping action of the TEC will vary 9.82 mW for every 1 mA change in $I_{TEC}$. As designed, the calorimeter has a theoretical current control resolution of 1.5 $\mu$A, therefore can maintain an accurate driving current which will vary according to the controller output.

The undesirable sensitivity of heat pumped to temperature differences across the TEC was addressed by controlling the calorimeter {\RT in a similar manner as shown in \cite{4314664}}. Instead of pumping heat through both TECs, only the left TEC is used to remove heat from the load, and the power is measured with the left heat flux sensor. The left TEC is controlled such that the heat flux through the {\RT right heat flux sensor} is zero, creating an adiabatic boundary on the right hand side of the HFL. {\RT In contrast to \cite{4314664}, one of the TECs is used to help regulate the temperature of the DUT and the other TEC is used to cool the apparatus.} Heat flux perturbations caused by the changing current command to the left TEC are eliminated through the use of a dead zone controller \cite{1566703}. The dead zone controller creates a small dead band around the reference input where the error is zero. In the case of the calorimeter, there is a $\pm$0.4 mW heat flux dead band around the reference value of 0 W for the right heat flux sensor.

\subsection{Taking a Measurement}
\begin{figure}
\centering
\includegraphics[width=\linewidth]{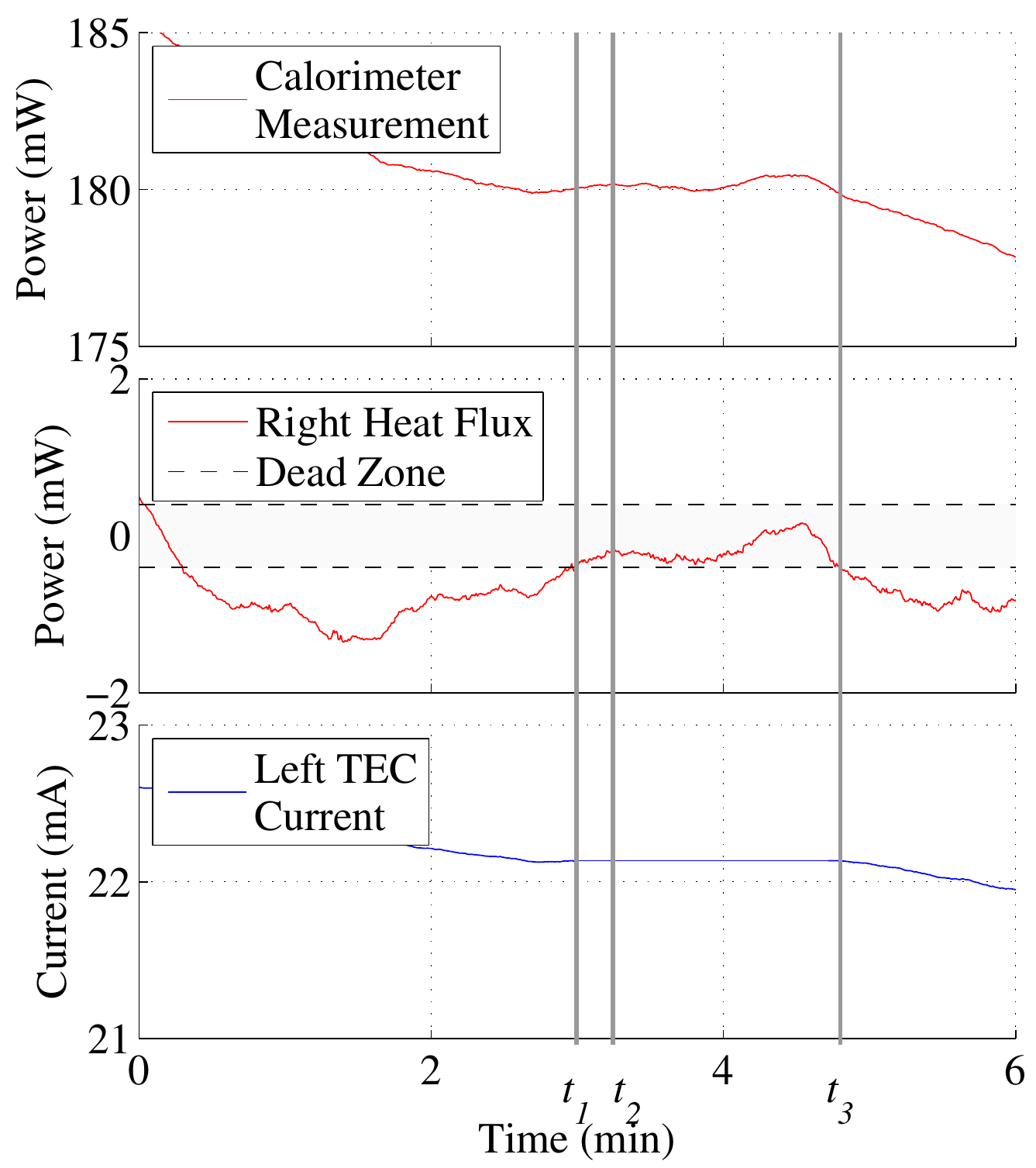}
\caption{Measurements over time from the calorimeter showing the dead zone implementation in the controller and the conditions to be met for a measurement to be taken. In this figure, the measurements between times $t_2$ and $t_3$ are valid calorimetric power readings.}
\label{fig:StabilityCriteria}
\end{figure}

The controller designed for the calorimeter is able to determine when the system has come to a thermal equilibrium and a reading can be taken. The process of taking a steady state heat flux measurement from the load (Fig.\ \ref{fig:StabilityCriteria}) is as follows:
\begin{enumerate}
\item Wait until the right heat flux measurement is within the dead zone, $t_1$, and start a 15 second timer.
\item Wait until the timer expires, at $t_2$. If right heat flux measurement is still within the dead zone, record a calorimetric power measurement. Continue recording a calorimetric measurement until the right heat flux measurement is no longer in the dead zone, at $t_3$.
\item The measurements between $t_2$ and $t_3$ are the power measurements from the calorimeter. 
\end{enumerate}

\section{Calibration}
\label{calibration}
The calorimeter was calibrated against the PPA at low frequency (135 Hz), where the PPA has a known error of $\pm$2.4 mW at 1 W and unity power factor\cite{PPAManual}. Due to the complexity of the heat leakage flow paths, {\RT such as those identified in the thermal simulations of Section \ref{sec:thermalsim}},  the number of measurement inputs that affect the estimated heat flux, and the requirement for as high a measurement accuracy as possible, an artificial neural network (ANN)  \cite{dayhoff1990neural} was used to obtain a functional mapping between all of the sensor inputs and the calibrated power measurement, as shown in \refeqn{eqn:caleqn}. In the literature, ANNs have been used successfully to calibrate sensors \cite{5303512, 5190072,293445,6616579} and scientific equipment \cite{4989939}.
\begin{equation}
P_{calorimeter} = f_{ANN}\left\{ m_1, m_2, \ldots ,m_7\right\} \label{eqn:caleqn}
\end{equation}
In this equation, $P_{calorimeter}$ is the power measurement of the calorimeter, $f_{ANN}$ is the functional mapping implemented by the ANN, and inputs $\left\{ m_1, m_2, \ldots ,m_7\right\}$ are given in Table \ref{tab:inputs}.

\begin{table}[h]    
\caption{Inputs to the ANN mapping function, $f_{ANN}$.}\label{tab:inputs}
	\begin{center}
		\begin{tabular}{ c | c}
		\hline
		$m_1$ & left heat flux \\
		$m_2$ & right heat flux \\
		$m_3$ & left PCB temp \\
		$m_4$ & right PCB temp \\
		$m_5$ & ambient temp \\
		$m_6$ & water temp \\
		$m_7$ & left TEC current 
		\end{tabular}
	\end{center}

\end{table}

\begin{figure}
\centering
\includegraphics[width=\linewidth]{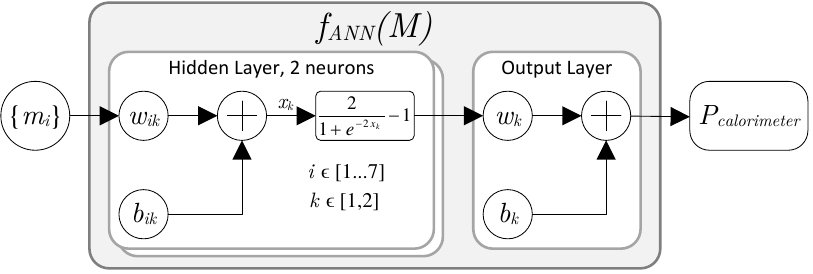}
\caption{Schematic diagram of the artificial neural network used to calibrate the calorimeter.}
\label{fig:ANN}
\end{figure}

A diagram of the neural network used in the calorimeter is shown in Fig.\ \ref{fig:ANN}. %
The high calorimeter speed made it possible to obtain a large amount of training data for the ANN. All six measurements shown in Fig.\ \ref{fig:CalControl} in addition to the current control signal to the left TEC formed the seven inputs into $f_{ANN}$. In order to avoid the common problem of over-fitting \cite{JMLR:v15:srivastava14a}, the smallest ANN structure that achieved a good fit was desired. Different sizes of ANN were evaluated by comparing their outputs to the low frequency power measurements of the PPA and calculating the average rms error for each set of measurements. The rms error is calculated with \refeqn{eqn:emeasrms}:

\begin{equation}
e_{meas-rms} = \sqrt{\left({P_{calorimeter} - P_{PPA}}\right)^2} \label{eqn:emeasrms}
\end{equation}

During initial testing, an ANN with a single neuron produced results with four times higher rms error than a 2 neuron network. Higher order networks (3 neurons or more) did not substantially improve the rms error. Therefore, a 2 neuron ANN was implemented.

\begin{figure}
\centering
\includegraphics[width=0.7\linewidth]{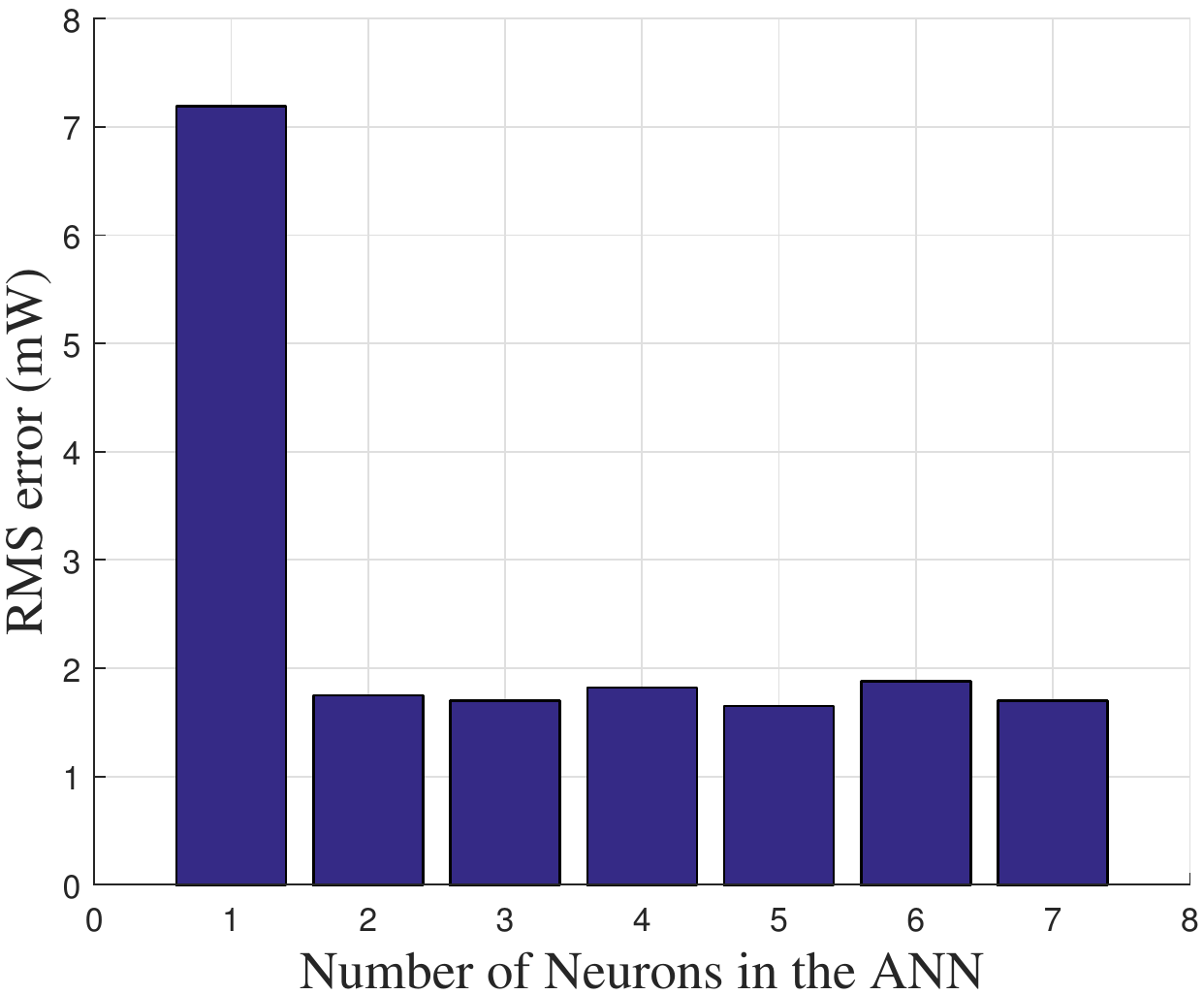}
\caption{RMS error of a 100 mW measurement using a different number of neurons in the ANN.}
\label{fig:NumberOfNeurons}
\end{figure}

The ANN was designed using MATLAB's neural network toolbox. The Levenberg-Marquardt back-propagation algorithm \cite{Levenberg} was used for training and the trained ANN was implemented in the calorimeter software, to provide real time calibrated power measurements.

\section{Experimental Results}

Two experiments were conducted with the calorimeter. Firstly the system was tested against the N4L PPA at low frequency in order to validate the ANN calibration technique. Secondly the calibrated calorimeter was used to test the high frequency power measurement accuracy of the N4L PPA.

\begin{figure}
\centering
\includegraphics[width=\linewidth]{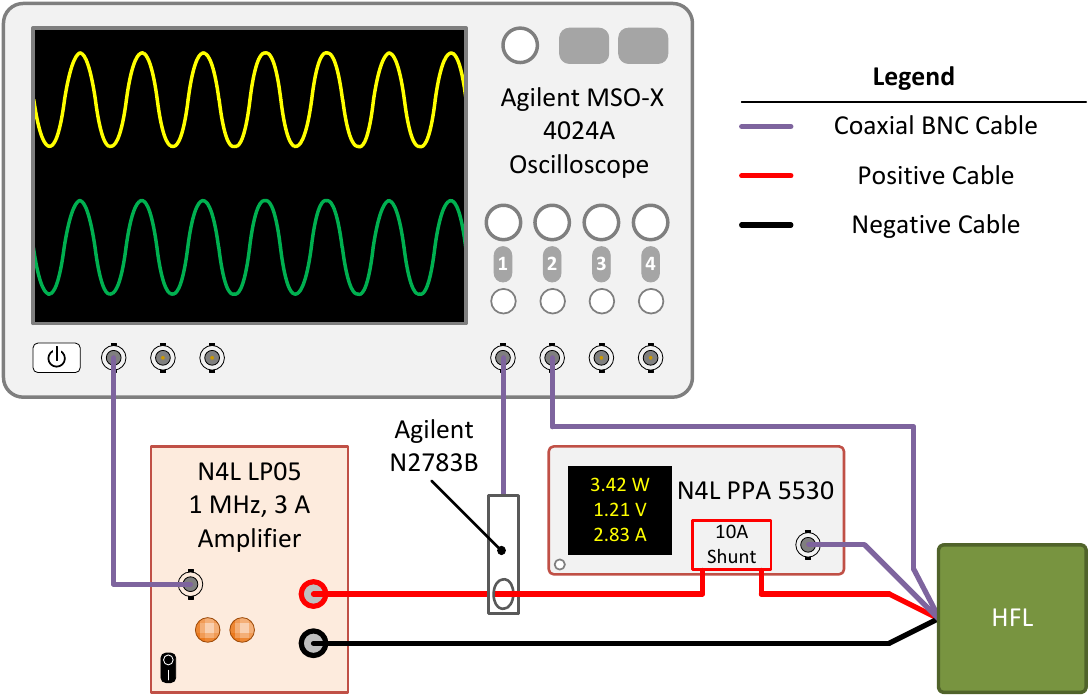}
\caption{{\RT Diagram of the experimental setup used to calibrate the Calorimeter, and then test the N4L PPA.}}
\label{fig:CalExperimentalSetup}

\end{figure}

{\RT The experimental setup is shown in Fig. \ref{fig:CalExperimentalSetup}. The HFL was driven by a high frequency amplifier, the N4L LPA05, that is capable of delivering AC current with a peak of 3 A at 1 MHz. A reference signal for the amplifier was provided by the built-in waveform generator of the Agilent MSO-X 4024A oscilloscope. Voltage and current measurements were taken with the oscilloscope and N4L PPA. An Agilent N2783B current probe was used to measure the current for the oscilloscope. The internal 10 A current shunt of the N4L PPA was used to measure the current for the PPA. Voltages for both pieces of equipment were measured with voltage probes.}

\subsection{Low frequency operation}

This section presents the results from the validation exercise at low frequency, showing the accuracy of the calorimeter as compared to the (calibrated) N4L PPA. The first step was to calibrate the calorimeter by recording several hours of data at a low frequency of 135 Hz where the N4L PPA has a known calibration. The calibration data was used to train the two neuron neural network as previously discussed.

After calibration, \emph{another independent} measurement set was taken across the entire power range at different power levels to the initial calibration power levels. This independent data set was not used to train the neural network, and was only used to demonstrate the accuracy of the calorimeter.

\begin{figure}
\centering
\includegraphics[width=\linewidth]{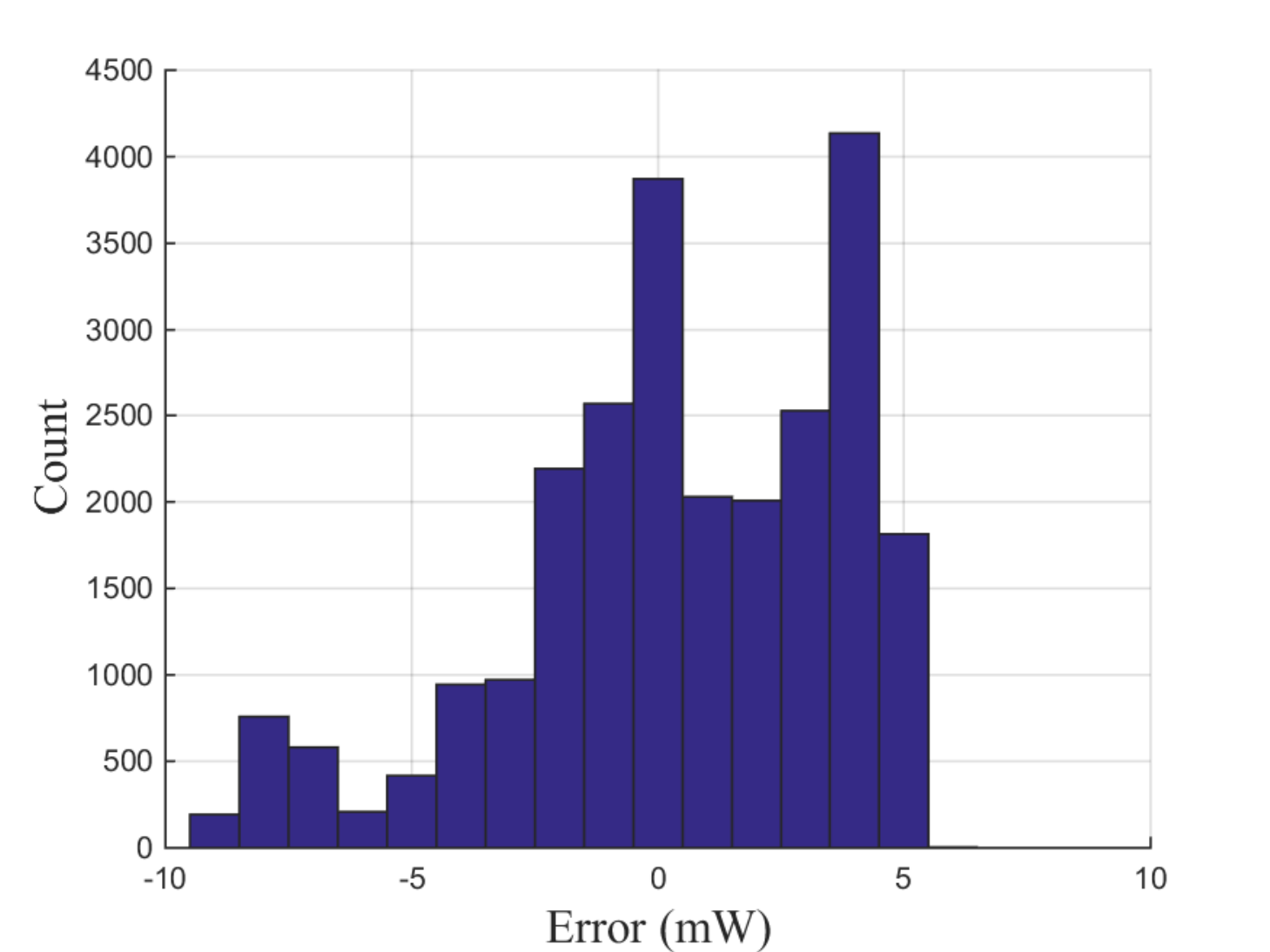}
\caption{Histogram of the calorimeter uncertainty over several measurements and power levels. The largest and smallest values were 5.5 mW and -8.9 mW respectively.}
\label{fig:ErrorHist}

\end{figure}

Analysis of this data showed the calorimeter power readings were no more than $\pm$ 9 mW away from the PPA measurement over the entire measurement range from 0.1 W to 5 W. Furthermore, 95\% of all the estimates were within $\pm$6 mW of the PPA measurements. Fig.\ \ref{fig:ErrorHist} shows a histogram of the errors between the `true' power measurements from the N4L PPA and the estimated measurements from the calibrated calorimeter. The worst case uncertainty of the N4L PPA during calibration was during the 5 W measurements and it was $\pm$ 5 mW\footnote{PPA uncertainty $= \left[0.03\% + 0.03\%/\textrm{PF} + \left(0.01\%\times\textrm{kHz}\right)/\textrm{PF}\right]\times\textrm{Reading} + 0.02\%\times\left(\textrm{VA Range}\right) = \left(0.03+0.03/0.9984+\left(0.01\times.135\right)/0.9984\right)/100\times5+0.02\times9/100 = 5$ mW; where PF is power factor.} \cite{PPAManual}. Therefore, the absolute uncertainty of the calorimeter within a 95 \% confidence interval is the root-sum-square of the uncertainties of the calorimeter and PPA, which yields $\pm$8 mW. 

The average amount of time required by the calorimeter to reach a steady measurement was 8.3 minutes. 

\begin{figure}
\centering
\includegraphics[width=0.8\linewidth]{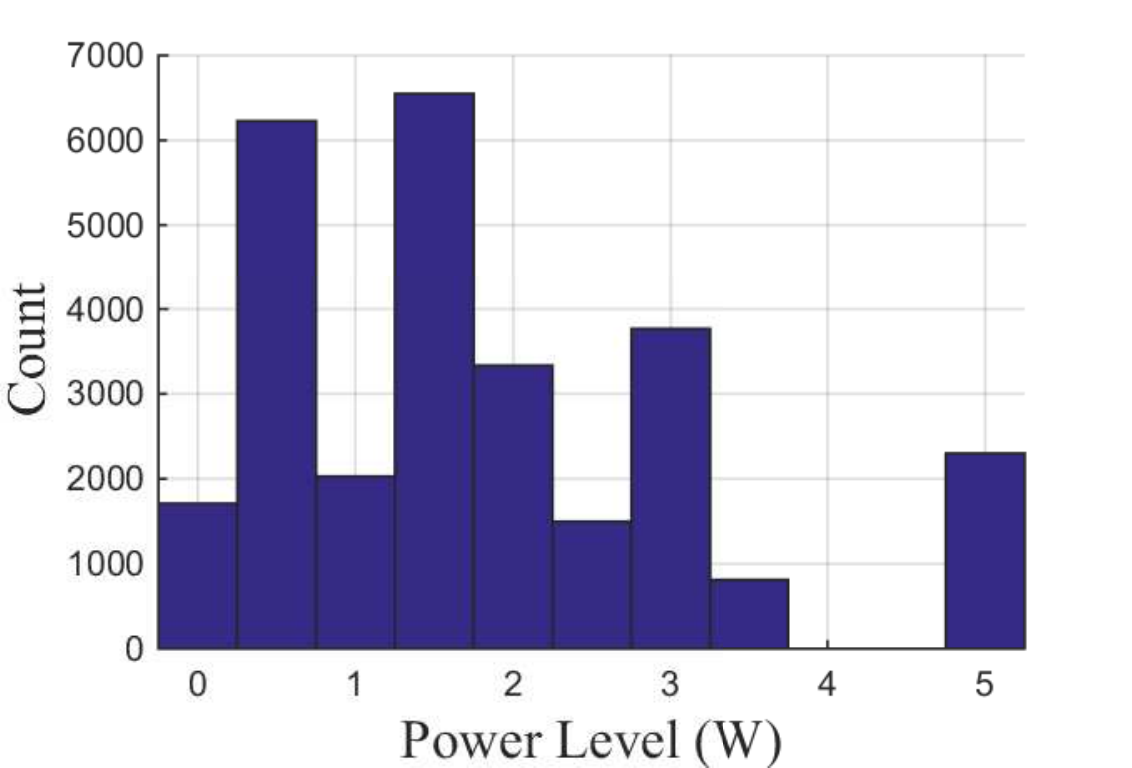}
\caption{{\RT Histogram of the number of measurements taken for different power levels to create the histogram of Fig.\ \ref{fig:ErrorHist}}}
\label{fig:MeasurementCounts}

\end{figure}

{\RT Fig.\ \ref{fig:MeasurementCounts} shows the number of counts for the different power levels tested used to generate the histogram of Fig.\ \ref{fig:ErrorHist}. We were interested in the performance of the calorimeter at the lower power levels, hence the discrepancy in counts between the higher power levels.} 

\subsection{High frequency operation}

During the high frequency tests, the nominal power delivered to the HFL was 1 W.

\begin{figure}
\centering
\includegraphics[width=\linewidth]{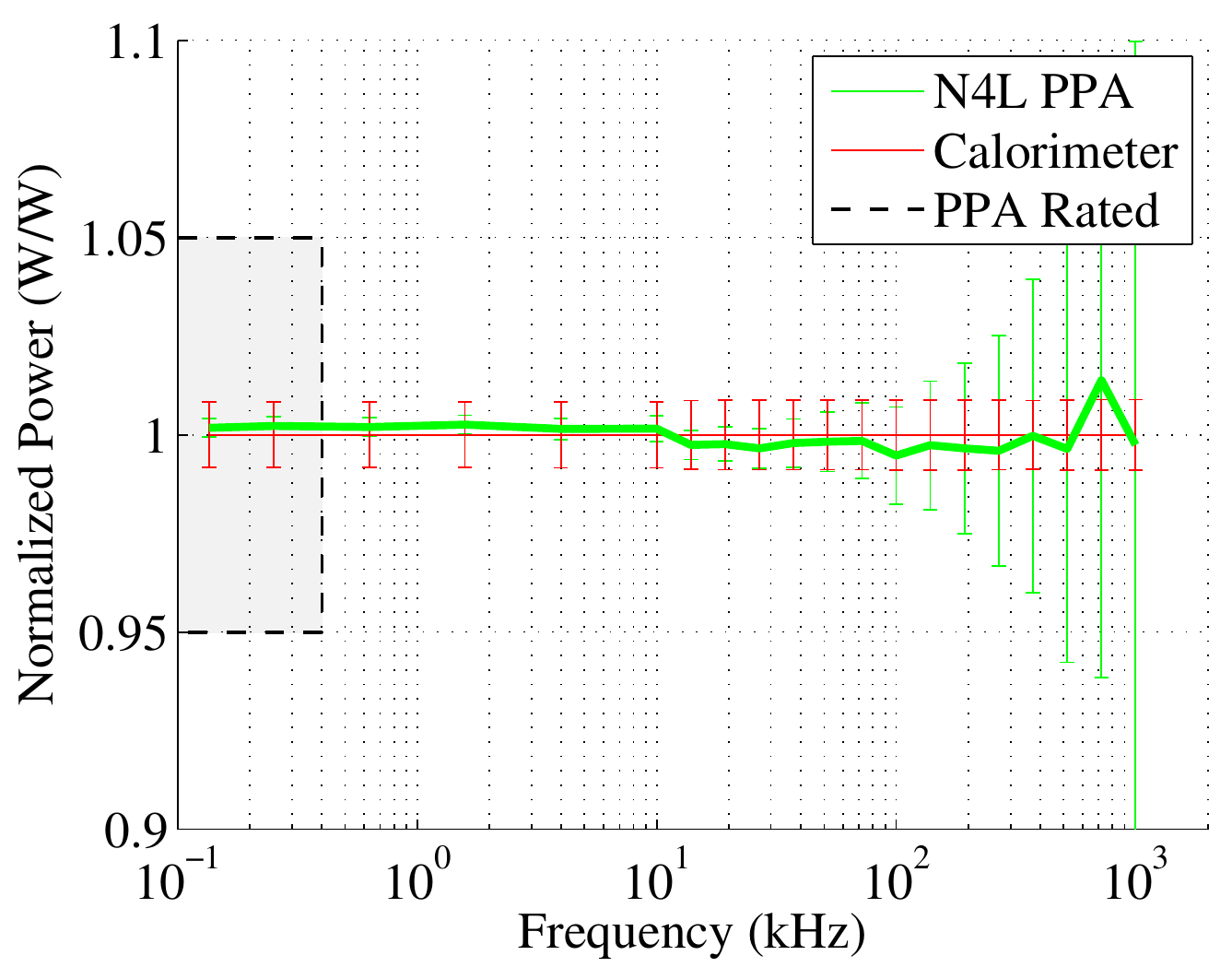}
\caption{Normalized values of power measurements from the PPA and Calorimeter across a range of frequencies at 1 W nominal. Normalized power measurements were used to display the data for ease of reading. All of the power measurements were taken between 1.00 W and 1.09 W. The shaded region is the region where N4L has rated their PPA. The error bars for the PPA outside of this region are an extrapolation of their error calculation found in the user manual \cite{PPAManual}.}
\label{fig:NormHFT}

\end{figure}

Fig.\ \ref{fig:NormHFT} shows the results of power measurements taken over a wide range of frequencies normalized to the calorimeter measurement at each frequency point.  The N4L PPA and calorimeter show excellent agreement throughout the frequency range. The error bars shown for the PPA are an extrapolation of the error calculation provided in the N4L PPA manual \cite{PPAManual}. N4L has rated the power measurement of the PPA5530 up to 400 Hz, depicted by the shaded region in Figure \ref{fig:NormHFT}.

\begin{figure}
\centering
\includegraphics[width=\linewidth]{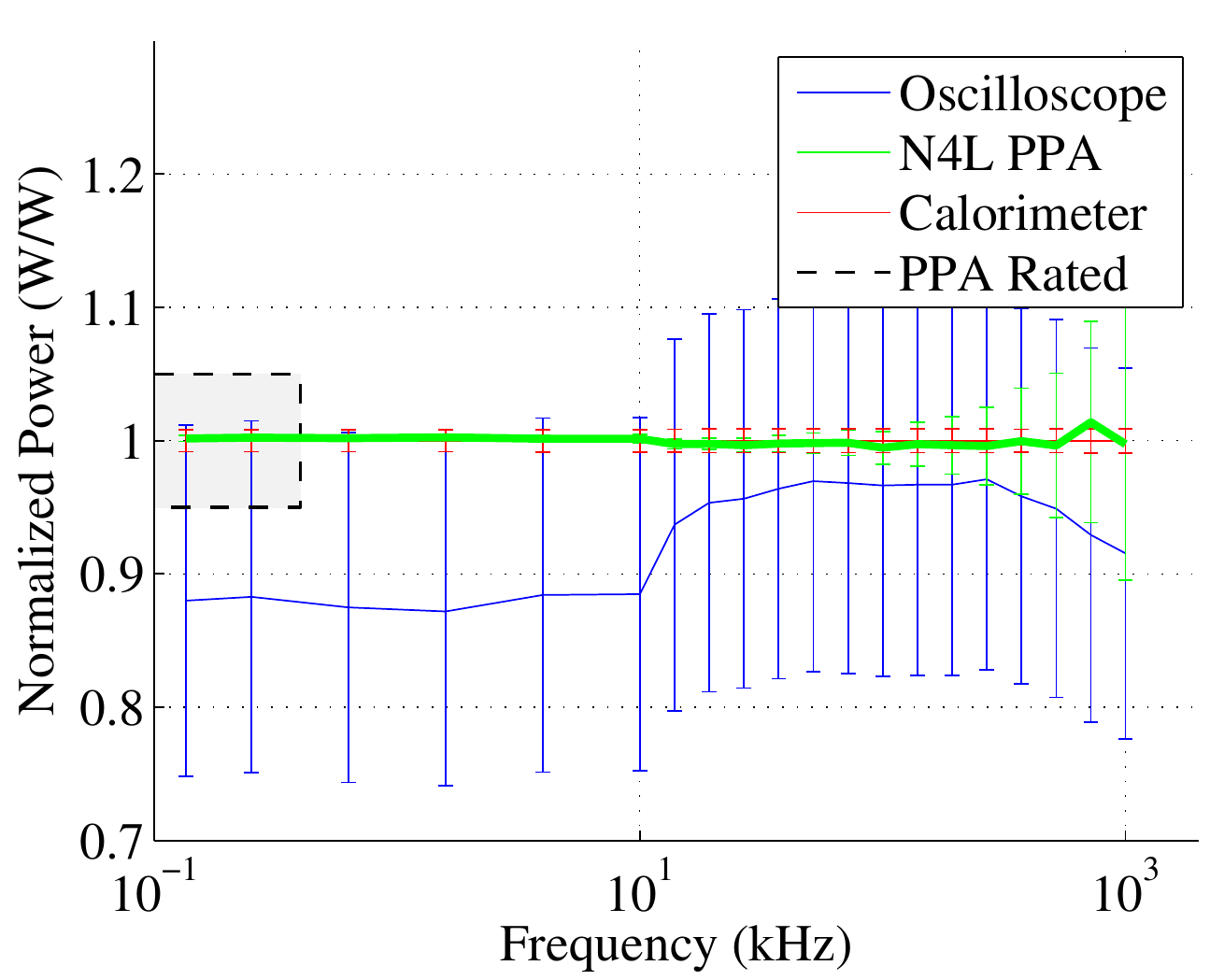}
\caption{Normalized values of power measurements from the PPA, Calorimeter and oscilloscope, using the same data as shown in Fig.\ \ref{fig:NormHFT}.}
\label{fig:NormHFTScope}

\end{figure}

The oscilloscope power measurements were less accurate as shown in Fig.\ \ref{fig:NormHFTScope}. However this is most likely due to the use of the Agilent N2783B current probe which has a full scale range of 50 A. The current measured in this experiment was about 1 A, and at this level the probe has an accuracy of $\pm$1.8\%\cite{N2780BCurrentProbeDatasheet}. Combined with the uncertainties of the oscilloscope \cite{MSOX4024ADatasheet}, the error in the measurement was an average of $\pm$ 144 mW over the entire frequency range.

\section{Conclusions}
We have presented a novel high speed calorimeter for calibrating PPAs at high frequencies, using Peltier elements and heat flux sensors to form a `sandwich' style system. An absolute accuracy of $\pm$ 8 mW with a 95\% confidence interval over a DUT power range of 0.1 W to 5 W and an average measurement time of 8.3 minutes was achieved. This was due to a number of innovations in our design: (1) the heat was removed from one side of the load cell only, whilst the other side was controlled to form an adiabatic boundary, (2) a dead zone controller was used to overcome the sensitivity of the Peltier elements to changes in the current through them and (3) an artificial neural network was used to calibrate the system.

The calorimeter was used to measure the accuracy of a precision power analyzer and an oscilloscope, finding that whilst the PPA demonstrated good accuracy to 1 MHz, the oscilloscope did not, most likely due to the current measurement probes being used at the bottom end of their full scale range. These results are relevant for those undertaking sensitive power measurements such as switching losses in power semiconductors at high frequencies, where care should be taken to ensure an accurate measurement is made.

%

\section*{ACKNOWLEDGMENT}
The authors acknowledge the financial support provided by Newtons4th Ltd.~(N4L), the Natural Sciences and Engineering Research Council of Canada (NSERC) and Jesus College Oxford. This work also benefited from equipment funded by the John Fell Oxford University Press (OUP) Research Fund. 


\bibliographystyle{IEEEtran}
\bibliography{references-frost}

\begin{IEEEbiography}[{\includegraphics[width=1in,height=1.25in,clip,keepaspectratio]{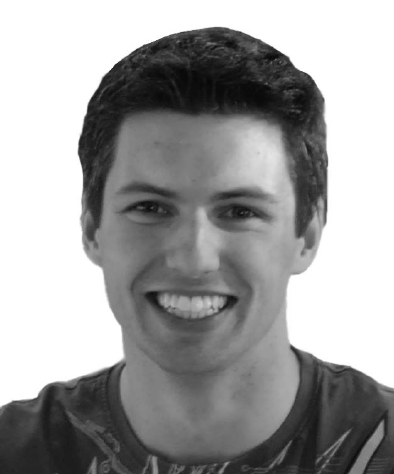}}]{Damien F. Frost}
received the B.A.Sc. and M.A.Sc. degrees from the University of Toronto, Canada in 2007 and 2009, respectively. Before starting his D.Phil (PhD) degree in the Energy and Power Group, Department of Engineering Science, University of Oxford, Oxford, U.K. in 2013, he worked in the solar industry as a co-founder and power electronics designer at ARDA Power. He is currently working towards his D.Phil at Oxford, where his research is focused on the application of power electronics and control theory to battery management systems.
\end{IEEEbiography}

\begin{IEEEbiography}[{\includegraphics[width=1in,height=1.25in,clip,keepaspectratio]{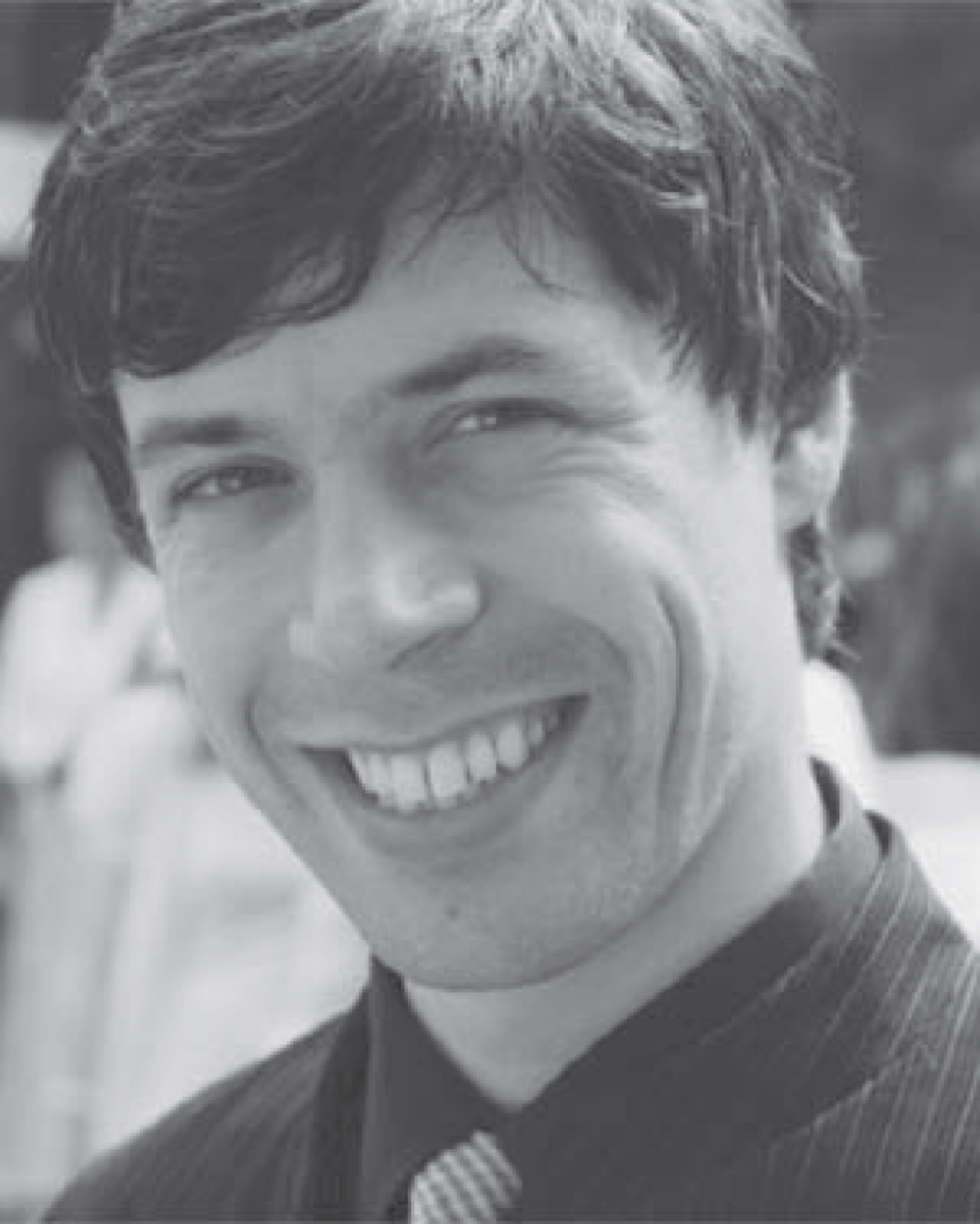}}]
{David A. Howey}
(M\textquoteright{}10) received the B.A. and M.Eng. degrees from Cambridge University, Cambridge, U.K., in 2002 and the Ph.D. degree from Imperial College London, London, U.K., in 2010. \RTT{He is currently an Associate Professor with the Energy and Power Group, Department of Engineering Science, University of Oxford, Oxford, U.K. His research is focused primarily on energy storage systems, including projects on model-based battery management, degradation, thermal management, and energy management for grid storage.}
\end{IEEEbiography}

\end{document}